\documentclass[prl,aps,twocolumn,showpacs,preprintnumbers,amsmath,amssymb]{revtex4}
\usepackage{epsfig}
\usepackage{color}
\usepackage{ifthen}
\usepackage{graphicx}
\usepackage{amsmath}
\usepackage{amssymb}

\usepackage{graphicx}
\usepackage{dcolumn}
\usepackage{bm}
\def\Dsl{\hbox{/\kern-.6700em\it D}} 
\def\dsl{\hbox{/\kern-.5300em$\partial$}}

\def\eqa{\begin{eqnarray}}
\def\eeqa{\end{eqnarray}}
\def\eq{\begin{equation}}
\def\eeq{\end{equation}}
\def\be{\begin{equation}}
\def\ee{\end{equation}}
\def\bea{\begin{eqnarray}}
\def\eea{\end{eqnarray}}

\newcommand{\dslash}{\not{\hbox{\kern-2pt $\partial$}}}
\newcommand{\pslash}{\not{\hbox{\kern-2.3pt $p$}}}
 \newtoks\nslashfraction
 \nslashfraction={.13}
 \newcommand{\nslash}[1]{\setbox0\hbox{$ #1 $}
   \setbox0\hbox to \the\nslashfraction\wd0{\hss \box0}/\box0 }

\begin{document}

\preprint{FERMILAB-PUB-08-072-A-T}

\title{Spin-Statistics Violations from Heterotic String Worldsheet Instantons}
\author{Mark G. Jackson}
\affiliation{Particle Astrophysics Center and Theory Group, Fermi National Accelerator Laboratory, Batavia IL 60510}

\date{\today}


\begin{abstract}
In this paper we consider the role that worldsheet instantons in the heterotic string could play in spin-statistics violations.  Such violations are nonperturbative in the string tension and so would not appear in the spacetime effective action, producing a unique signature of string theory and the details of compactification.  By performing a Bogomol'nyi transformation  it is shown that there are no instanton solutions in the simplest model proposed by Harvey and Liu, but it is conjectured that more sophisticated models may yield solutions.  If such instantons do exist, their effect might be measured by upcoming experiments.
\end{abstract}

\maketitle

\section{Introduction}
Ever since its discovery, the Aharonov-Bohm Effect \cite{Aharonov:1959fk} has fascinated physicists for its quantum-mechanical implications, which are completely foreign to our classical prejudices.  It also has also proved useful in producing fractional statistics for charged particles in 2+1 dimensions \cite{Wilczek:1982wy}.  As the particle completes a circuit around a localized magnetic flux core as in Fig. 1(a), its wavefunction will pick up a phase, altering the naive expectation that bosons (fermions) must always be in symmetric (antisymmetric) wavefunctions.  Similarly, in 3+1 dimensions, a coupling of the form $\int B \wedge dA$ would produce the same type of effect, whereby a particle (coupled to $A_\mu$) passing through a closed loop (coupled to $B_{\mu \nu}$) acquires a statistical phase \cite{Aneziris:1990gm} \cite{Almeida:2001nt}, as shown in Fig. 1(b).  

An instantonlike mechanism to utilize this fact in Heterotic superstring theory  \cite{Gross:1984dd} (where such a $BF$ coupling arises naturally from anomaly cancellation) was proposed by Harvey and Liu \cite{Harvey:1990wa}, whereby one string will momentarily open up and pass over another string before collapsing again, as shown in Figure 2.  The magnitude of this spin-statistics-violating effect was estimated to be of order $e^{-1/\alpha' E^2}$, assuming that one string must open up to at least the de Broglie wavelength of the other.  Naively, $1/\sqrt{\alpha'} \sim 10^{16}$ GeV, and so this is prohibitively too small to be observed, but if $1/\sqrt{\alpha'} \sim$ TeV (as in some recent warped models \cite{Randall:1999ee} \cite{Kachru:2003sx}) then perhaps this effect is observable at achievable energies and worth revisiting.  Note that this intrinsically stringy effect would never show up in the spacetime effective action, which is a perturbative expansion in small $\alpha'$.  

In this paper, I attempt to explicitly construct these worldsheet instantons but find there are no solutions in the model proposed by Harvey and Liu.  I then consider additional terms that may yield a solution but are considerably more difficult to analyze.  These solutions, if they exist, would likely scale not with the energy but rather with fixed parameters in the theory, making them easier to detect with current experiments at low energy.

\section{The Instanton Action}
We will assume that this instanton process happens in 3+1 Minkowski space after compactification.  The  action for the first string with momentum $k_1$ and coupled to the Kalb-Ramond 2-form $B$ is
\[ S_{1} = \frac{1}{2 \pi \alpha'} \int d^2 z \left[ \partial X^\mu {\bar \partial} X^\nu (\delta_{\mu \nu} + 2 \pi \alpha' B_{\mu \nu}) + 2 \pi \alpha'  \delta^2(z, {\bar z}) k_1\cdot X  \right]. \]
Note that the term containing $B$ is imaginary and thus produces a phase in the path integral, and that we are considering worldsheet instanton solutions so the momentum $k_1$ is real.  The action for the second string (which we approximate as a particle) with momentum $k_2$ coupled with charge $q$ to the pseudoanomalous $U(1)$ gauge field $A$ is
\[ S_2 = \int dl \ \left[  \frac{1}{2 \alpha'} {\dot Y} \cdot {\dot Y} + {\dot Y} \cdot \left( iq A-k_2 \right) \right]. \]
Again note that the term coupling to $A$ is imaginary.  The spacetime action governing the gauge fields $F=dA$ and ${\tilde H}=dB-A \wedge dA$ is
\[ S_{gauge} = \int d^4 x \left[ \frac{3 \alpha'}{32g^2} {\tilde H}^2 + \frac{1}{4g^2} F ^2 \right]  + \theta \int B \wedge F  \]
where $g^2$ and $\theta$ are the dimensionless effective 4D couplings after compactification.  This spacetime action is introduced as 
\[ \int [ \mathcal D A] [\mathcal D B] \ e^{iS_{gauge}} \]
meaning the imaginary string source terms for $B$ and $A$ are real source terms from the perspective of the spacetime action. The worldsheet and worldline then produce, respectively, $F$ and ${\tilde H}$ flux tubes with width $\sim \sqrt{\alpha'} /  \theta g^2$ (this is reversed from the usual case due to the $\theta B \wedge F$ term).  If we approximate these as infinitesimally thin, we may neglect the gauge kinetic terms and integrate the fields out, resulting in the effective action equal to\begin{equation}
\label{linking}
 S_{eff} \sim - \frac{i q}{\theta} \frac{\epsilon^{\mu \nu \rho  \lambda }}{4 \pi^2} \int d\Sigma_{\mu \nu}(X) \int dY_\rho \frac{ (X-Y)_\lambda}{|X-Y|^4}.
 \end{equation}
Thus the phase shift of the wavefunction will be proportional to this ``linking number,"  a topological quantity equal to the number of times the worldline $Y$ will pass through the worldsheet $\Sigma(X)$.  We will now attempt to construct solutions whereby this happens dynamically.
\begin{figure}
\begin{center}
\includegraphics[width=3.2in]{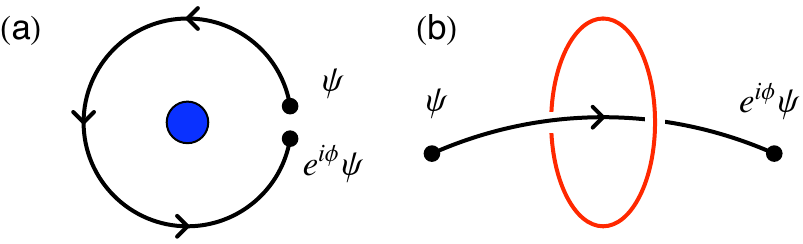}
\caption{(a) In 2+1 dimensions a charged particle's wavefunction will acquire a phase after a circuit around a flux tube, (b) A similar phase can be acquired in 3+1 dimensions for a particle passing through a loop.}
\end{center}
\end{figure}

\section{BPS Transformation}
In the heterotic string theory, with different compactifications we can get different values of $\theta = c/32 \pi^2$, where $c$ is determined by the massless fermion content of the theory.  In the case of compactification on a Calabi-Yau manifold \cite{Candelas:1985en} we break $SO(32) \rightarrow SU(3) \times SO(26) \times U(1)$ and then embed the spin connection in the gauge group.  This yields $c=-\frac{3}{2} \chi$, where $\chi$ is the Euler number of the Calabi-Yau, and the fermion charges are $q=\pm 1, \pm 2$.

Rather than simply look for generic solutions to the equations of motion, we look for solutions which minimize the action for a given value of the linking number.  To facilitate this, let us repeat the previous derivation of the phase shift more explicitly.  By taking the strong coupling limit $\theta g^2 \rightarrow \infty$ for fixed $\theta$ as above, we can neglect the gauge kinetic terms and so easily obtain exact solutions
\begin{eqnarray*}
B_{\mu \nu}(x) &=& \frac{q \epsilon_{\mu \nu \rho \lambda}}{\theta} \int dl \ \partial^{[ \rho} G(x-Y) {\dot Y}^{\lambda]} , \\
A_{\mu} (x) &=& \frac{i \epsilon_{\mu \nu \rho \lambda} }{2 \theta} \int d^2 z \ \partial^{[\nu} G(x-X) \partial X^{\rho} {\bar \partial} X^{\lambda ]}
\end{eqnarray*} 
where $G(x-y)$ is the 4D Green's function
\[ G(x-y) = - \frac{1}{2 \pi^2} \frac{1}{|x-y|^2} . \]
Substituting these back into the action produces a worldsheet path integral with topological phase factor $\theta \int B \wedge F$, which is proportional to the linking number $N = \epsilon^{\mu \nu \rho \lambda} \int d \Sigma_{\mu \nu}(X) \int dY_\rho \partial_\lambda G(X-Y)$,
\begin{equation}
\label{wspi}
 \int [ \mathcal D X] [ \mathcal D Y] \ e^{-\frac{1}{2 \pi \alpha'} \int d^2 z | \partial (X - \alpha' k_1 \ln |z|)|^2 - \frac{1}{2 \alpha'} \int dl |\dot Y - \alpha' k_2|^2 - i \frac{qN}{\theta} }. 
 \end{equation}

The path integrals over $X$ and $Y$ will select trajectories such that the strings will take the least action path to produce the linking.  For the zero-modes $x_0,y_0$ this is trivial, as can be seen by momentarily considering the first string to also be pointlike.  By choosing $x_0=y_0=0$, the strings will both simply travel along straight lines in the direction of their respective momenta until they intersect at $X(z,{\bar z})=Y(l)=0$, then a linking is obtained by briefly expanding the first string's worldsheet.  Any other choice of $x_0$ or $y_0$ would require the strings to either ``swerve"  or else further enlarge the worldsheet, both of which increase the action.  We can then perform a Bogomol'nyi transformation and express the (real part of the) action as a sum of squares plus a topological term:
\begin{eqnarray*}
S &=&  \frac{1}{2 \pi \alpha'} \int d^2 z \ \left| \partial (X^\mu - \alpha' k_1^\mu \ln |z|) \right. \\
&& \left. + \ i \frac{ \pi qC\alpha' }{\theta} {\epsilon^\mu}_{ \nu \rho \lambda} \partial (X^\nu + \alpha' k_1^\nu \ln |z|) \int dY^\rho \partial^\lambda G(X-Y) \right|^2 \\
&+& \frac{1}{2 \alpha'} \int dl \ |\dot Y - \alpha' k_2|^2 + \frac{CqN}{\theta}.
\end{eqnarray*}
The constant $C$ is yet to be determined. 

\begin{figure}
\begin{center}
\includegraphics[width=1.5in]{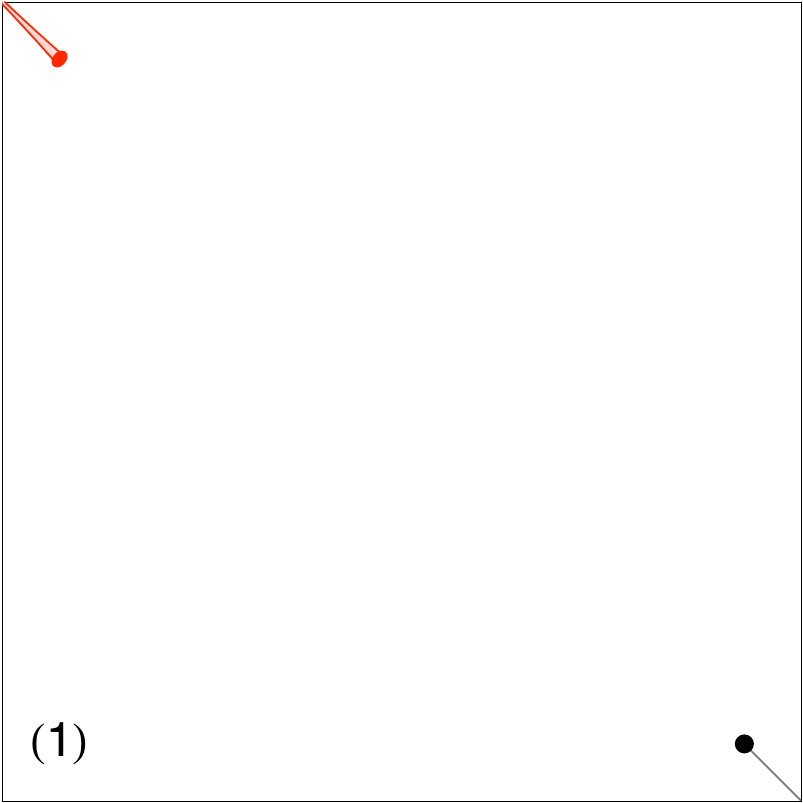}
\includegraphics[width=1.5in]{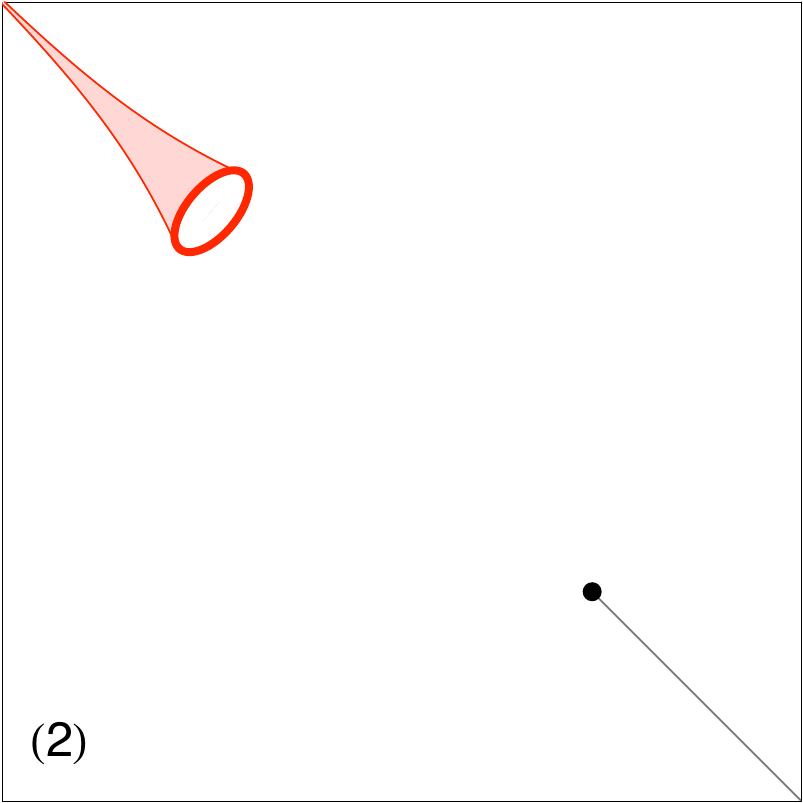}
\includegraphics[width=1.5in]{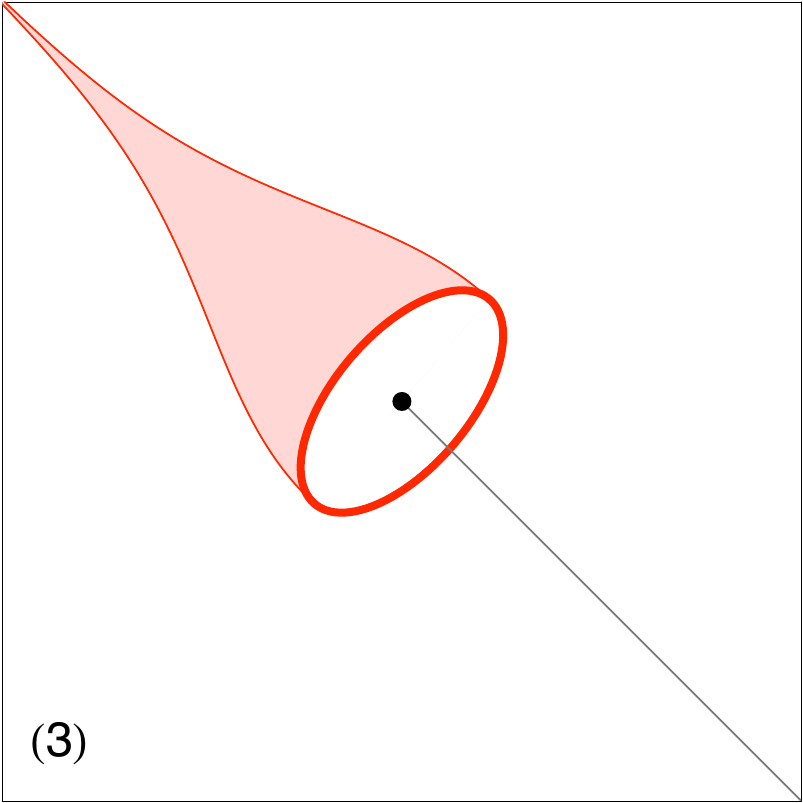}
\includegraphics[width=1.5in]{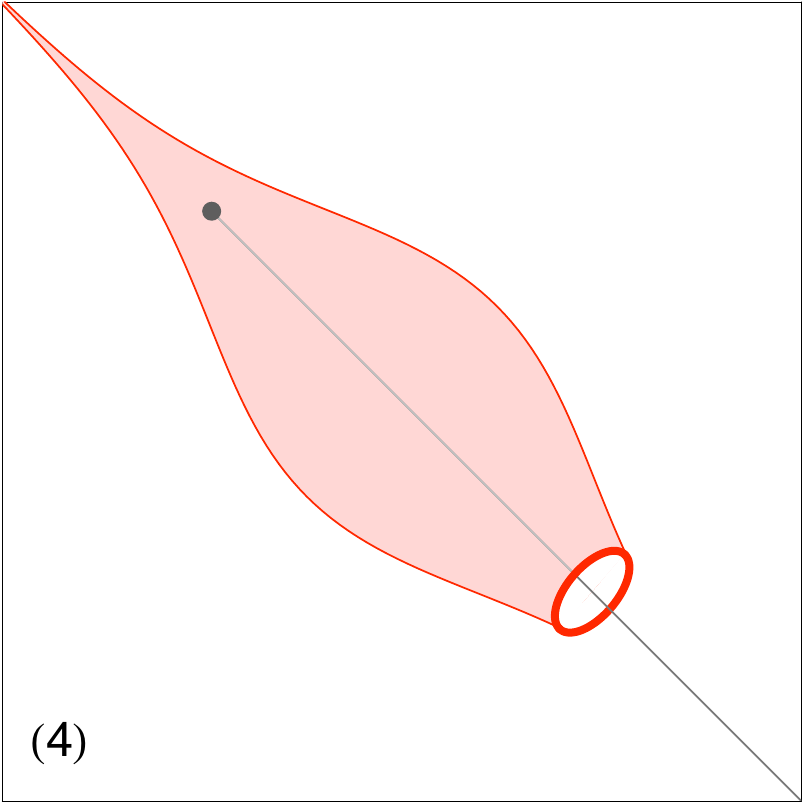}
\caption{Worldsheet instantonlike process whereby one string momentarily expands sufficiently to envelop another, producing a phase in the string path integral.}
\end{center}
\end{figure}

A minimal-action solution is then obtained by setting the two squared terms to zero.  The one for $Y$ is trivial and yields $Y(l) = k_2 l$ as expected.  To solve the one for $X$, first integrate the Green's function over $Y$,
\begin{eqnarray*}
 \int \ dY^\mu \int \frac{d^4 p}{(2 \pi)^4} \frac{e^{ip \cdot [X-Y(l)] }}{p^2} &=& \alpha' k_2^\mu \int \frac{d^4 p}{(2 \pi)^3} \frac{e^{ip \cdot X} }{p^2} \delta \left( \alpha' k_2 \cdot p \right) \\
 &=& - \frac{ {\hat k}_2^\mu}{4 \pi |X_\perp |}
 \end{eqnarray*}
 where $X_\perp$ is the component of $X$ transverse to $k_2$.  Since both strings couple to gauge fields they cannot be massless and so $X_\perp$ is spacelike.  The BPS equation for $X(z,{\bar z})$ can then be written as
\begin{eqnarray}
\nonumber
z \partial X^\mu  &=& \alpha'  \left( {\delta^\mu}_\nu + i \frac{ qC \alpha'}{4 \theta} {\epsilon^\mu}_{ \nu \rho \lambda}  \frac{ X_\perp^\rho {\hat k}_2^\lambda}{ |X_\perp |^3} \right)^{-1} \\
\label{bpsx}
&& \hspace{0.4in} \cdot \left( {\delta^\nu}_\gamma - i \frac{ qC \alpha'}{4 \theta} {\epsilon^\nu}_{ \gamma \kappa \sigma}  \frac{ X_\perp^\kappa {\hat k}_2^\sigma}{ |X_\perp |^3} \right) k_1^\gamma.
\end{eqnarray}
To solve this we decompose $X$ into a complete basis, beginning with (timelike) ${\hat k}_2$ and then defining the three spacelike unit vectors ${\hat x}, {\hat y}, {\hat z}$ as transverse to this, such that ${\hat x}, {\hat y}$ are also transverse to ${\hat k}_1$.  The ansatz then consists of a complex function $f$ (${\bar f}$) representing the positive (negative) chirality in the ${\hat x}-{\hat y}$ plane and a real function $h$ in the ${\hat z}$-direction,
\begin{eqnarray}
\label{x}
X(z,{\bar z}) &=& \alpha' (k_1 \cdot {\hat k}_2) {\hat k}_2 \ln |z| \\
\nonumber
&& \hspace{-0.5in} + f(z,{\bar z}) \left( \frac{{\hat x}-i{\hat y}}{2} \right) + {\bar f}(z,{\bar z}) \left( \frac{{\hat x}+i{\hat y}}{2} \right) + h(z,{\bar z}) {\hat z} . 
\end{eqnarray}
Then $|X_\perp|^2 = |f|^2 + h^2$.  Choosing the convention $\epsilon_{ {\hat k}_2 {\hat x} {\hat y} {\hat z}} = 1$, we can substitute (\ref{x}) into (\ref{bpsx}) to see that the ${\hat k}_2$ component is trivially satisfied, whereas $f$ and $h$ must obey the relations
\begin{eqnarray}
\nonumber
z \partial f &=& Nf \left[  \frac{ qC \alpha' h}{4 \theta |X_\perp|^3} + 1 \right], \\ 
\nonumber
z \partial {\bar f} &=&  N {\bar f} \left[  \frac{ qC \alpha' h}{4 \theta |X_\perp|^3}  - 1 \right], \\ 
\label{bpsfh}
z \partial h &=& \alpha' k_1 \cdot {\hat z} \left( \frac{1 +  \left( \frac{ qC \alpha'}{4 \theta |X_\perp|^3} \right)^2(|f|^2-h^2) }{ 1 - \left( \frac{ qC \alpha' }{4 \theta |X_\perp|^3} \right)^2 \left( |f|^2+h^2 \right) } \right)
\end{eqnarray}
where the (suggestively named) real function $N(z,{\bar z})$ is 
\[ N = \frac{ \alpha' k_1 \cdot {\hat z} \left( \frac{ qC \alpha'}{2 \theta |X_\perp|^3} \right)}{ 1 - \left( \frac{ qC \alpha' }{4 \theta |X_\perp|^3} \right)^2 \left( |f|^2+h^2 \right) } . \]
We now switch to $(\tau, \sigma)$ coordinates such that $z = e^{\tau + i \sigma}$.  Inspection of the real and imaginary parts of the equations in (\ref{bpsfh}) implies that $h=h(\tau)$, $N$ is a constant, and $f$ is of the form
\[ f(\tau,\sigma) = f_0 \exp \left[ N \left( \int d \tau \ \frac{ qC \alpha' h(\tau)}{4 \theta |X_\perp|^3}+ i \sigma \right) \right] .  \]
This identifies $N$ as the linking number.  For $N$ to be constant places an algebraic constraint on $|X_\perp|$,
\[ \frac{4 \theta (|f|^2+h^2)^{3/2} }{ qC \alpha'} =  \frac{\alpha' k_1 \cdot {\hat z}}{N} \pm \sqrt{ \left( \frac{\alpha' k_1 \cdot {\hat z}}{N} \right)^2 + |f|^2+h^2}. \]
This is inconsistent except in the trivial cases $f = h = 0$ or $C=N=0$.  Therefore no instantons exist in this formalism.

\section{Discussion and Possible Resolutions}
The lack of instanton solutions should have been anticipated from the action in (\ref{wspi}), which shows that there is no interaction between the two strings, merely a phase.  With nothing to set a lower bound on how close the strings may approach, all paths are smoothly contractable to the trivial solution.  The intuition of \cite{Harvey:1990wa} that a distance cutoff is fixed via the de Broglie wavelength fails because the linking number projects onto the transverse worldsheet-worldline separation and so is not aware of their respective momentum eigenstates.

One may try to remedy the situation by relaxing the assumption of infinite coupling and instead consider a finite value of $\theta g^2$.  Such an approach would modify the Green's function to give the flux tubes widths and so might seem to provide a distance cutoff $\Delta x \sim \sqrt{\alpha '} / \theta g^2 $.  Unfortunately this would still not produce an interaction term in the worldsheet action, merely change the action's topological term from the linking number into a finite-width-version of the linking number.

The most natural way to produce a valid interaction is to recall that the (left-moving component of the) first string may also carry a charge $Q$ under the pseudoanomalous $U(1)$ gauge field, 
\begin{equation}
\label{ds1}
 \Delta S_1 =  \frac{1}{2 \pi} \int  d^2 z \ J(z) A_\mu {\bar \partial} X^\mu 
 \end{equation}
where $J$ is the holomorphic $U(1)$ current normalized so that $\oint dz \ J(z) = 2 \pi i Q$.  Were we to also approximate this string as infinitesimally thin, a Kaluza-Klein reduction of the worldsheet would result in the purely imaginary term 
\[ \frac{1}{2 \pi} \int  d^2 z \ J(z) A_\mu {\bar \partial} X^\mu \approx i Q \int d \tau A_\mu {\dot X}^\mu \]
just like the analogous term in $S_2$, which contributes only a path integral phase, but at nonzero string size this also contains a real component, which affects the worldsheet dynamics.  At infinite coupling $\theta g^2 \rightarrow \infty$ this additional term produces only worldsheet self-interactions, so we must go to finite coupling to yield an interaction between the two strings.  The maximal radius $R$ of the worldsheet should be where the tension of the string $\sim 1/ \alpha' $ is balanced by the force of electrostatic repulsion $\sim g^2 qQ / R^2$, so there is an equilibrium reached at roughly 
\[ R \sim \sqrt{ g^2 q Q \alpha' } . \]
The addition of (\ref{ds1}) to the action for strings with $q Q > 0$ could then plausibly produce instanton solutions, and could be analyzed using techniques similar to those employed here.  Unfortunately, explicit solutions for this model are likely much more difficult to construct due to the necessity of finite coupling.

\section{Experimental Bounds}
If such instantons did exist, they would produce phases in the path integral leading to small violations of spin-statistics, most noticeably the Pauli Exclusion Principle (PEP) for fermions.  Depending on whether they scaled with energy or fixed parameters in the theory, observing these violations could either come from high-energy or high-precision experiments.  Energy scales of order 14 TeV will soon be available from the Large Hadron Collider, though it is uncertain to what extent it would be sensitive to extremely small phases in scattering amplitudes.  The first precision test of the PEP was performed by Ramberg and Snow \cite{Ramberg:1988iu} by running current through a copper cylinder and looking for forbidden X-ray transitions.  A similar technique used by the ongoing VIP (VIolations of the Pauli exclusion principle) Experiment \cite{VIP} has thus far constrained the deviation away from Fermi statistics in terms of the Ignatiev-Kuzmin-Greenberg-Mohapatra $\beta$ parameter \cite{Ignatiev:1987cd} \cite{Greenberg:1988um} as
\[ \frac{\beta^2}{2} \leq 4.5 \times 10^{-28}. \]
This bound is expected to improve another 2 orders of magnitude over the next few years due to larger integrated currents.  Though the energy scale is low at only 8 keV, the incredible precision means this might be a viable way of detecting  superstring-motivated violations.

\section{Conclusion}
In this paper we have examined heterotic string worldsheet instantons, which could potentially produce a statistical phase and violate spin-statistics.   They are found to not exist in the simplest case considered here but more general models may produce solutions.  Unfortunately the analysis of such generalizations is beyond the scope of this paper, requiring explicit solutions to gauge theories at finite coupling.  In the auspicious case in which there is a measurable effect, this could be an experimentally viable way of testing string theory and could provide detailed information about the microscopic parameters.  It would also be interesting to include the effects of extra dimensions rather than a simple compactification.  The violation of spin-statistics, even if slightly, could have dramatic physical and even cosmological \cite{Jackson:2007tn} consequences.
\section{Acknowledgements} 
I would like to thank Jeff Harvey, Joe Lykken, Catalina Petrascu, and David Tong for useful discussions. This work was supported by the DOE at Fermilab.


\begin{thebibliography}{99}

\small
\parskip=0pt plus 2pt

\bibitem{Aharonov:1959fk}
  Y.~Aharonov and D.~Bohm,
  ``Significance of electromagnetic potentials in the quantum theory,''
  Phys.\ Rev.\  {\bf 115}, 485 (1959).
  
\bibitem{Wilczek:1982wy}
  F.~Wilczek,
  ``Quantum Mechanics Of Fractional Spin Particles,''
  Phys.\ Rev.\ Lett.\  {\bf 49}, 957 (1982).

\bibitem{Aneziris:1990gm}
  C.~Aneziris, A.~P.~Balachandran, L.~Kauffman and A.~M.~Srivastava,
  ``Novel Statistics for Strings and String `Chern-Simons' Terms,''
  Int.\ J.\ Mod.\ Phys.\  A {\bf 6}, 2519 (1991).
  
\bibitem{Almeida:2001nt}
  C.~A.~S.~Almeida,
  ``Remarks on topological models and fractional statistics,''
  Braz.\ J.\ Phys.\  {\bf 31}, 277 (2001)
  [arXiv:hep-th/0105232].
  
\bibitem{Gross:1984dd}
  D.~J.~Gross, J.~A.~Harvey, E.~J.~Martinec and R.~Rohm,
  ``The Heterotic String,''
  Phys.\ Rev.\ Lett.\  {\bf 54}, 502 (1985).
  
\bibitem{Harvey:1990wa}
  J.~A.~Harvey and J.~Liu,
  ``Strings and Statistics,''
  Phys.\ Lett.\  B {\bf 240}, 369 (1990).
  
\bibitem{Randall:1999ee}
  L.~Randall and R.~Sundrum,
  ``A large mass hierarchy from a small extra dimension,''
  Phys.\ Rev.\ Lett.\  {\bf 83}, 3370 (1999)
  [arXiv:hep-ph/9905221].
  
\bibitem{Kachru:2003sx}
  S.~Kachru, R.~Kallosh, A.~Linde, J.~M.~Maldacena, L.~McAllister and S.~P.~Trivedi,
  ``Towards inflation in string theory,''
  JCAP {\bf 0310}, 013 (2003)
  [arXiv:hep-th/0308055].
  
   %
\bibitem{Candelas:1985en}
  P.~Candelas, G.~T.~Horowitz, A.~Strominger and E.~Witten,
  ``Vacuum Configurations For Superstrings,''
  Nucl.\ Phys.\  B {\bf 258}, 46 (1985).

    \bibitem{Ramberg:1988iu}
  E.~Ramberg and G.~A.~Snow,
  ``A new experimental limit on small violation of the Pauli principle,''
  Phys.\ Lett.\  B {\bf 238}, 438 (1990).
  
\bibitem{VIP}
  S.~Bartalucci {\it et al.} [The VIP Collaboration],
  ``New experimental limit on Pauli Exclusion Principle violation by electrons,''
  Phys.\ Lett.\  B {\bf 641}, 18 (2006)
    [arXiv:quant-ph/0612116].
  
\bibitem{Ignatiev:1987cd}
  A.~Y.~Ignatiev and V.~A.~Kuzmin,
  ``Is Small Violation Of The Pauli Principle Possible?,''
   Yad.\ Fiz.\  {\bf 46}, 786 (1987).
  
\bibitem{Greenberg:1988um}
  O.~W.~Greenberg and R.~N.~Mohapatra,
  ``Phenomenology of Small Violations of Fermi and Bose Statistics,''
  Phys.\ Rev.\  D {\bf 39}, 2032 (1989).
    
\bibitem{Jackson:2007tn}
  M.~G.~Jackson and C.~J.~Hogan,
  ``A new spin on quantum gravity,''
Int.\ J.\ Mod.\ Phys.\  D {\bf 17}, 567 (2008)
  [arXiv:hep-th/0703133].

\end{thebibliography}
\end{document}